\documentclass{emulateapj}

\usepackage{natbib}
\usepackage{amsmath}

\newcommand{\about}{$\sim\!\!$~}
\newcommand{\kms}{\,km\,s$^{-1}$}

\def\lsim{\hbox{\rlap{\raise 0.425ex\hbox{$<$}}\lower 0.65ex\hbox{$\sim$}}}
\def\gsim{\hbox{\rlap{\raise 0.425ex\hbox{$>$}}\lower 0.65ex\hbox{$\sim$}}}

\def\arcsec{\hbox{$^{\prime\prime}$}}
\def\arcdeg{\mbox{$^\circ$}}

\newcommand{\hst}{\protect\hbox{$H\!ST$} }

\newcommand{\vsi}{\protect\hbox{$v_{\rm Si~II}$}}
\newcommand{\vsiz}{\protect\hbox{$v_{\rm Si~II}^{0}$}}

\usepackage{color}

\shorttitle{SN~2011iv: Maximum-Light UVOIR Spectrum}
\shortauthors{Foley et~al.}

\begin{document}

 \title{The First Maximum-Light Ultraviolet through Near-Infrared Spectrum of a Type I\lowercase{a} Supernova\altaffilmark{\hst,\mag}}

\def\hst{1}
\def\mag{2}
\def\cfa{3}
\def\clay{4}
\def\mpia{5}
\def\ab{6}
\def\aar{7}
\def\stock{8}
\def\berk{9}
\def\ipmu{10}
\def\car{11}
\def\li{12}
\def\wurz{13}
\def\mit{14}
\def\unc{15}

\author{
{Ryan~J.~Foley}\altaffilmark{\cfa,\clay},
{Markus~Kromer}\altaffilmark{\mpia},
{G.~Howie~Marion}\altaffilmark{\cfa},
{Giuliano~Pignata}\altaffilmark{\ab},
{Maximilian~D.~Stritzinger}\altaffilmark{\aar,\stock},
{Stefan~Taubenberger}\altaffilmark{\mpia},
{Peter~Challis}\altaffilmark{\cfa},
{Alexei~V.~Filippenko}\altaffilmark{\berk},
{Gast\'{o}n~Folatelli}\altaffilmark{\ipmu},\\
{Wolfgang~Hillebrandt}\altaffilmark{\mpia},
{Eric~Y.~Hsiao}\altaffilmark{\car},
{Robert~P.~Kirshner}\altaffilmark{\cfa},
{Weidong~Li}\altaffilmark{\berk,\li},\\
{Nidia~I.~Morrell}\altaffilmark{\car},
{Friedrich~K.~R\"{o}pke}\altaffilmark{\mpia,\wurz},
{Franco~Ciaraldi-Schoolmann}\altaffilmark{\mpia},\\
{Ivo~R. Seitenzahl}\altaffilmark{\mpia,\wurz},
{Jeffrey~M.~Silverman}\altaffilmark{\berk},
{Robert~A.~Simcoe}\altaffilmark{\mit},\\
{Zachory~K.~Berta}\altaffilmark{\cfa},
{Kevin~M.~Ivarsen}\altaffilmark{\unc},
{Elisabeth~R.~Newton}\altaffilmark{\cfa},\\
{Melissa~C.~Nysewander}\altaffilmark{\unc}, and
{Daniel~E.~Reichart}\altaffilmark{\unc}
}

\altaffiltext{\hst}{
Based on observations made with the NASA/ESA Hubble Space Telescope,
obtained at the Space Telescope Science Institute, which is operated
by the Association of Universities for Research in Astronomy, Inc.,
under NASA contract NAS 5-26555. These observations are associated
with program GO-12592.
}
\altaffiltext{\mag}{
This paper includes data gathered with the 6.5~m Magellan Telescopes
located at Las Campanas Observatory, Chile.
}
\altaffiltext{\cfa}{
Harvard-Smithsonian Center for Astrophysics,
60 Garden Street, 
Cambridge, MA 02138, USA
}
\altaffiltext{\clay}{
Clay Fellow. Electronic address rfoley@cfa.harvard.edu .
}
\altaffiltext{\mpia}{
Max-Planck-Institut f\"{u}r Astrophysik,
Karl-Schwarzschild-Strasse 1,
D-85748 Garching bei M\"{u}nchen, Germany
}
\altaffiltext{\ab}{
Departamento de Ciencias Fisicas,
Universidad Andres Bello,
Avda. Republica 252,
Santiago, Chile
}
\altaffiltext{\aar}{
Department of Physics and Astronomy,
Aarhus University,
Ny Munkegade,
DK-8000 Aarhus C, Denmark
}
\altaffiltext{\stock}{
The Oskar Klein Centre,
Department of Astronomy,
Stockholm University,
AlbaNova,
10691 Stockholm, Sweden
}
\altaffiltext{\berk}{
Department of Astronomy,
University of California,
Berkeley, CA 94720-3411, USA
}
\altaffiltext{\ipmu}{
Kavli Institute for the Physics and Mathematics of the Universe,
Todai Institutes for Advanced Study,
the University of Tokyo,
Kashiwa, Japan 277-8583 (Kavli IPMU, WPI)
}
\altaffiltext{\car}{
Carnegie Observatories,
Las Campanas Observatory,
La Serena, Chile 
}
\altaffiltext{\li}{
Deceased 12 December 2011
}
\altaffiltext{\wurz}{
Institut f\"{u}r Theoretische Physik und Astrophysik, 
Universit\"{a}t W\"{u}rzburg, 
Am Hubland, 
D-97074 W\"{u}rzburg, Germany
}
\altaffiltext{\mit}{
MIT-Kavli Institute for Astrophysics and Space Research
77 Massachusetts Ave \#37-664D
Cambridge, MA 02139, USA
}
\altaffiltext{\unc}{
Department of Physics and Astronomy,
University of North Carolina at Chapel Hill,
Chapel Hill, NC, USA
}

\defcitealias{Kirshner93}{K93}

\begin{abstract}
We present the first maximum-light ultraviolet (UV) through
near-infrared (NIR) Type Ia supernova (SN~Ia) spectrum.  This spectrum
of SN~2011iv was obtained nearly simultaneously by the \textit{Hubble
Space Telescope} at UV/optical wavelengths and the Magellan Baade
telescope at NIR wavelengths.  These data provide the opportunity to
examine the entire maximum-light SN~Ia spectral-energy distribution.
Since the UV region of a SN~Ia spectrum is extremely sensitive to the
composition of the outer layers of the explosion, which are
transparent at longer wavelengths, this unprecedented spectrum can
provide strong constraints on the composition of the SN ejecta, and
similarly the SN explosion and progenitor system.  SN~2011iv is
spectroscopically normal, but has a relatively fast decline ($\Delta
m_{15} (B) = 1.69 \pm 0.05$~mag).  We compare SN~2011iv to other
SNe~Ia with UV spectra near maximum light and examine trends between
UV spectral properties, light-curve shape, and ejecta velocity.  We
tentatively find that SNe with similar light-curve shapes but
different ejecta velocities have similar UV spectra, while those with
similar ejecta velocities but different light-curve shapes have very
different UV spectra.  Through a comparison with explosion models, we
find that both a solar-metallicity W7 and a zero-metallicity
delayed-detonation model provide a reasonable fit to the spectrum of
SN~2011iv from the UV to the NIR.
\end{abstract}



\section{Introduction}\label{s:intro}

Type Ia supernovae (SNe~Ia) are the energetic end of C/O white dwarfs
in binary systems that reach temperatures and densities high enough to
cause a runaway chain of nuclear reactions \citep[for a review,
see][]{Hillebrandt00}.  Despite having collected observations of
thousands of SNe~Ia, we still do not know what progenitor systems
produce SNe~Ia and do not fully understand how the star explodes.
Measurements of SNe~Ia led to the discovery of the accelerating
universe \citep{Riess98:Lambda, Perlmutter99}, and observing SNe~Ia is
still one of the best ways to constrain cosmological parameters.  It
behooves us to understand the physics behind SN~Ia explosions, and
with additional knowledge of the progenitors and explosions, we hope
to further improve the utility of SNe~Ia for measuring cosmic
distances.

SN~Ia UV spectra are dominated by a forest of overlapping lines from
iron-group elements (IGEs).  UV photons are repeatedly absorbed and
re-emitted in those lines and gradually scattered redward where lower
opacities allow them to escape.  The UV is crucial to the formation of
the optical spectral-energy distribution (SED) of SNe~Ia
\citep{Sauer08}, and extremely sensitive to both the progenitor
composition and explosion mechanism \citep{Hoflich98, Lentz01}.
Observations of the UV directly probe the composition of the outermost
layers of ejecta, which are transparent at optical wavelengths soon
after explosion.

After light-curve shape correction \citep[e.g.,][]{Phillips93}, the
luminosity of a SN~Ia still depends significantly on host-galaxy
environment \citep{Kelly10, Lampeitl10:host, Sullivan10}.  This may
indicate that there are subtle environmental effects imprinted on the
progenitor stars that directly affect our luminosity calibration.
Different progenitor metallicity may affect the outcome of the
explosion \citep{Timmes03}, the relation between luminosity and
light-curve shape \citep{Howell07, Mazzali01, Mazzali06}, and the
observed UV spectrum especially at $\lambda < 2600$~\AA\
\citep{Hoflich98, Lentz00}.

At optical wavelengths SNe~Ia have remarkably uniform luminosity ($\sigma
\approx 0.16$~mag; e.g., \citealt{Hicken09:lc}), after correcting for
light-curve shape and color. This relationship extends to the $U$
band, but with larger scatter \citep{Jha06:lc}.  The scatter can be
further reduced to 0.11~mag after making a correction based on a
measurement of the ejecta velocity \citep{Foley11:vel, Foley11:vgrad,
Foley12:vel}.  The intrinsic $B-V$ color of SNe~Ia correlates strongly
with ejecta velocity, with redder SNe having higher velocity.  This is
explained as additional line blanketing in the $B$ band of the
high-velocity SNe, and this trend should extend to the UV
\citep{Foley11:vel}.

High-quality UV spectra for low-$z$ SNe~Ia are of particular importance
to the calibration of high-$z$ SNe~Ia.  Since the observed SN light
originates as rest-frame UV, most $z > 1$ SNe~Ia only have rest-frame
UV light curves, and low-$z$ UV spectra are critical for understanding
these data.  Our current lack of UV spectra limits K-corrections and
even SN classification at $z > 1$
\citep{Riess07}.

SNe~Ia bright enough for the \textit{International Ultraviolet
Explorer} were rare \citep{Foley08:uv}, and only one SN~Ia has a
published high signal-to-noise ratio (S/N) \textit{Hubble Space
Telescope} (\textit{HST}) spectrum near maximum light covering
wavelengths $\lesssim 2900$~\AA\ -- SN~1992A \citep[hereafter,
K93]{Kirshner93}.  \textit{Swift} has obtained spectra of several
SNe~Ia, but except for a few SNe \citep[e.g.,
SN~2009ig;][]{Foley12:09ig}, the spectra are generally of such low S/N
that detailed spectral analysis is not feasible \citep{Bufano09}.  A
Cycle 13 \textit{HST} program (GO-10182; PI Filippenko) attempted to
observe several SNe~Ia in the UV; unfortunately, the STIS spectrograph
failed before any data could be obtained.  The program was executed
using the ACS prism, which does not provide sufficient resolution to
distinguish spectral features \citep{Wang12}.  In Cycle 17, 30 SN~Ia
were observed at a single epoch by \textit{HST} (GO-11721; PI Ellis);
however, these spectra did not probe below 2900~\AA\ \citep{Cooke11}.
It has been 20 years since the last high-S/N \textit{true}-UV
maximum-light SN~Ia spectrum has been observed and published.

SN~2011iv was discovered at an unfiltered magnitude of 12.8 on 2011
December 2.57 (UT dates are used throughout) by \citet{Drescher11}.
There was no object detected on 2011 October 29.57 to a limit of
18.8~mag.  It was discovered in NGC~1404, an elliptical galaxy in the
Fornax cluster at $cz = 1947$~\kms\ \citep{Graham98} that also hosted
SN~2007on \citep[e.g.,][]{Stritzinger11} and is at $D = 20.4$~Mpc
($\mu = 31.53 \pm 0.07$~mag\footnote{This measurement is consistent
with the \citet{Blakeslee10} weighted average distance modulus for the
Fornax cluster of $\mu = 31.54 \pm 0.02$~mag, but is larger than other
distance estimates that have higher uncertainty.} from a surface
brightness fluctuation measurement; \citealt{Blakeslee10}).

\citet{Chen11} and \citet{Stritzinger11:11iv} obtained optical spectra
of SN~2011iv on 2011 December 3.7 and 4.1, respectively, and
determined that it was a young SN~Ia.  We triggered multiple programs
to study the photometric and spectroscopic evolution of the SN, its
circumstellar environment, its polarization, its energetics, and other
aspects.  In particular, we triggered our \textit{HST}
target-of-opportunity program to obtain UV spectra of SNe~Ia
(GO-12592; PI Foley; \citealt{Foley11:11iv}).  The first of seven UV
spectra was obtained on 2011 December 11.08 and covered wavelengths of
1615--10,230~\AA.  Nearly simultaneously (2011 December 11.24), we
obtained a near-infrared (NIR) spectrum with the Folded Port Infrared
Echellette (FIRE) spectrograph \citep{Simcoe08, Simcoe10} on the 6.5~m
Magellan Baade Telescope.  In this \textit{Letter}, we focus on this
first UV-optical-NIR (UVOIR) spectrum spanning 0.16--2.5~$\mu$m.



\section{Observations}\label{s:obs}

\begin{figure}
\begin{center}
\epsscale{1.1}
\rotatebox{0}{
\plotone{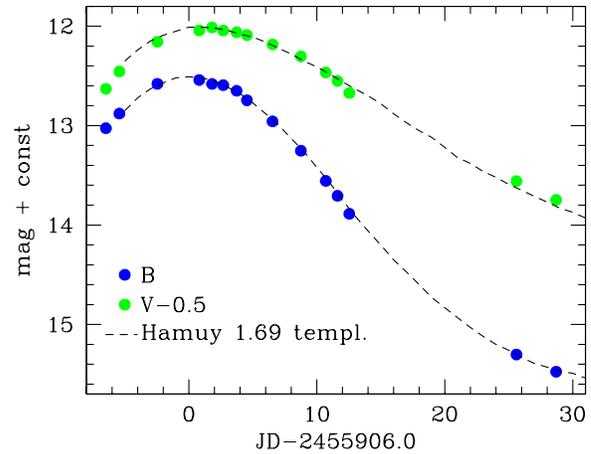}}
\caption{$B$ (blue points) and $V$ (green points) light curves of
SN~2011iv.  The dashed curves correspond to the $\Delta m_{15} =
1.69$~mag \citet{Hamuy96:temp} light-curve templates.}\label{f:lc}
\end{center}
\end{figure}

Optical photometry of SN~2011iv was collected with the Panchromatic
Robotic Optical Monitoring and Polarimetry Telescopes (PROMPT) 3 and 5
\citep{Reichart05}.  Basic data reduction (bias and flat-field
correction) was performed using standard routines in IRAF.  Local
standard stars were measured by \citet{Stritzinger11}.  The complex
background around SN~2011iv was modeled with a low-order polynomial
surface.  Once the galaxy contamination was removed, photometry was
performed using the point-spread-function fitting technique.  We
present our $B$ and $V$ light curves in Figure~\ref{f:lc}.

SN~2011iv was observed by \textit{HST} using the STIS spectrograph on
2011 December 11.08, corresponding to $t = 0.6$~days relative to $B$
maximum (see Section~\ref{s:results}).  The observations were obtained
over two orbits with three different gratings, all with the $52\arcsec
\times 0.\arcsec2$ slit.  Exposures of 2200 and 1350~s utilized the
near-UV MAMA detector and the G230L grating.  Two exposures of 100~s
were taken with both the CCD/G430L and CCD/G750L setups, respectively.
The three setups yield a combined wavelength range of
1615--10,230~\AA.

The data were reduced using the standard \textit{HST} Space Telescope
Science Data Analysis System (STSDAS) routines to bias subtract,
flat-field, extract, wavelength-calibrate, and flux-calibrate each SN
spectrum.

Almost immediately after obtaining the STIS spectrum (on 2011 December
11.24), we obtained a NIR spectrum (0.83--2.5~$\mu$m) of SN~2011iv
with FIRE.  SN~2011iv was observed in the high-resolution echellette
mode with the 0.\arcsec6 slit, yielding a resolution of $R = 6000$ or
about 50~\kms.  Four frames were taken on source with 240~s exposures
using ABBA nodding.

The data were reduced using a custom-developed IDL pipeline
(FIREHOSE), which evolved from the MASE pipeline used for optical
echelle reductions \citep{Bochanski09}.  The NIR sky flux was modeled
using off-source pixels as described by \citet{Kelson03} and
subtracted from the science frame before extraction, which was
weighted by a simple boxcar profile.  An A0V star was observed for
telluric corrections, following the procedures outlined by
\citet{Vacca03}. The \texttt{xtellcor} procedure, derived from the
Spextool package \citep{Cushing04}, performs both telluric correction
and relative flux calibration. The corrected echelle orders are then
combined into a single one-dimensional spectrum.

\begin{figure*}
\begin{center}
\epsscale{1.1}
\rotatebox{0}{
\plotone{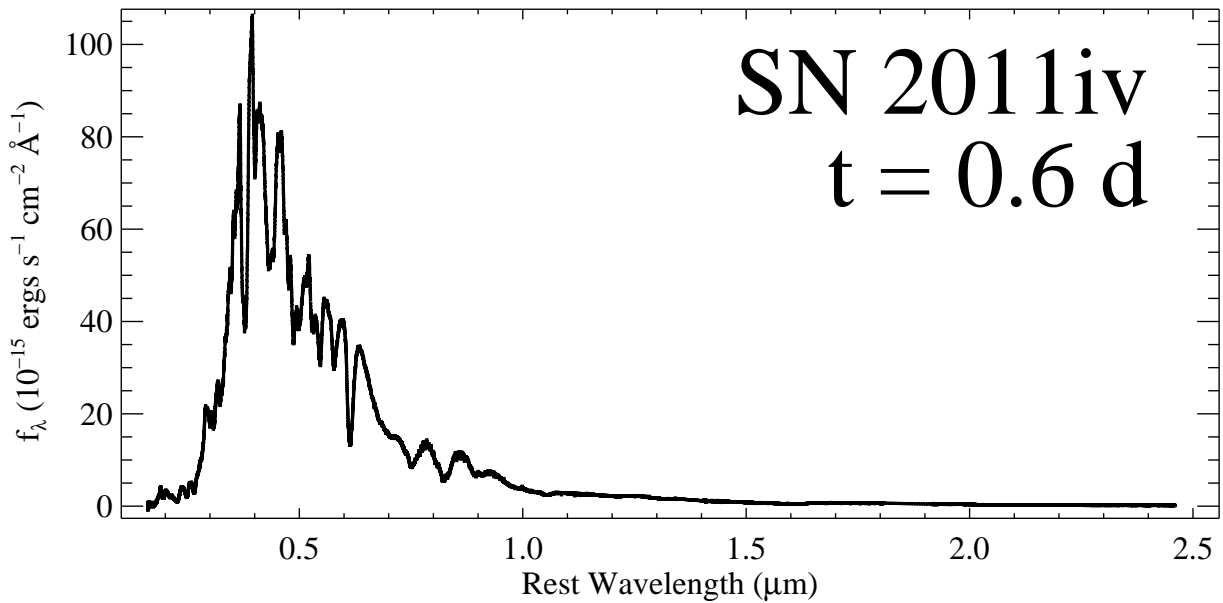}}
\caption{UVOIR \textit{HST}/STIS and Magellan/FIRE maximum-light spectrum
of SN~2011iv.}\label{f:spec}
\end{center}
\end{figure*}

\begin{figure*}
\begin{center}
\epsscale{1.1}
\rotatebox{0}{
\plotone{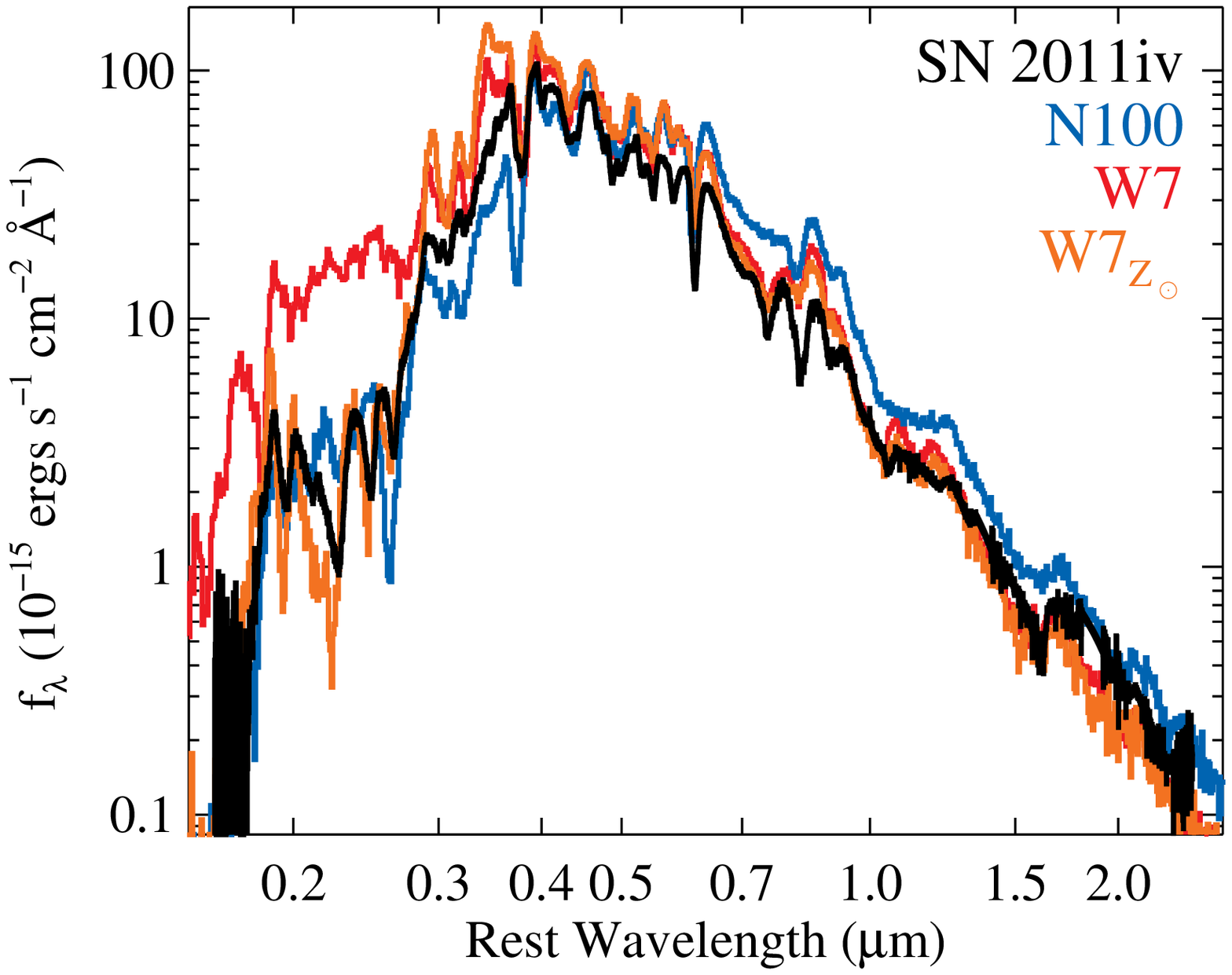}}
\caption{UVOIR maximum-light spectrum of SN~2011iv (black curve).
This spectrum is the same as that shown in Figure~\ref{f:spec}, but
with a different scale to highlight the UV and NIR portions.  The
blue, red, and orange curves represent model spectra generated from a
zero-metallicity delayed-detonation N100 model, the W7 model, and the
solar-metallicity polluted W7$_{Z_{\sun}}$ model,
respectively.}\label{f:model}
\end{center}
\end{figure*}

There is substantial wavelength overlap between the STIS and FIRE
spectra (8295--10,230~\AA).  Using this overlap region, we scaled the
FIRE spectrum to the STIS spectrum and combined the two spectra.  The
combined spectrum (1615--24,880~\AA) is presented in
Figures~\ref{f:spec} and \ref{f:model}; it has been corrected for
Milky Way reddening of $E(B-V) = 0.011$~mag
\citep{Schlegel98}.

\section{Results}\label{s:results}

Our spectrum of SN~2011iv is the first high-S/N, near-maximum, true-UV
SN~Ia spectrum since that of SN~1992A \citepalias{Kirshner93}; the
earliest high-S/N UV SN~Ia spectrum yet published; and the first
contemporaneous and continuous UVOIR spectrum of a SN~Ia.  This unique
dataset provides the opportunity to examine the full SED, producing
strong constraints on SN~Ia models.

Fitting the $B$ and $V$ light curves (Figure~\ref{f:lc}) with a
third-order polynomial function, we estimate that SN~2011iv reached
maximum light in the $B$ band on 2,455,$906.0 \pm 0.3$ JD and peaked
at $B = 12.53 \pm 0.03$~mag and $V = 12.51 \pm 0.03$~mag.  The
$B_{\max}-V_{\max}$ pseudo-color is consistent with negligible
reddening in the host galaxy.  Assuming a Milky Way extinction of
$A_{V} = 0.038$~mag \citep{Schlegel98} and our preferred distance
modulus, SN~2011iv peaked at $M_{V} = -19.06 \pm 0.08$~mag. The
optical light curves of SN~2011iv indicate that it is a relatively
fast-declining SN with $\Delta m_{15} (B) = 1.69 \pm 0.05$~mag; the
light curves are well matched by the $\Delta m_{15} = 1.69$~mag
\citet{Hamuy96:temp} templates (Figure~\ref{f:lc}).

As seen in Figure~\ref{f:spec}, optical emission dominates the SEDs of
SNe~Ia at maximum light.  The median flux over 100~\AA\ bins centered
at 2500, 3000, and 3500~\AA\ is 2.4, 18.1, and 50.0\% that of the peak
flux (in $f_{\lambda}$ units), respectively.  Similarly, the median
flux over 500~\AA\ bins centered at 1, 1.5, and 2~$\mu$m is 3.6, 0.8,
and 0.4\% that of the peak flux (in $f_{\lambda}$ units),
respectively.

SN~2011iv has a relatively strong silicon ratio, $\mathcal{R}({\rm
Si~II})$ \citep{Nugent95}, of $0.50 \pm 0.05$.  This measurement is
consistent with its decline rate.  Optical spectra of SN~2011iv are
extremely similar to those of SNe~1992A \citepalias{Kirshner93} and
2004eo \citep{Pastorello07:04eo}.  Both SNe~1992A and 2004eo also had
relatively fast decline rates ($\Delta m_{15} (B) = 1.48$ and
1.46~mag, respectively), but were still spectroscopically ``normal.''

Using the method of \citet{Blondin06}, we determine the velocity of
the \ion{Si}{2} $\lambda 6355$ feature to be $v_{\rm Si~II} = -10,630
\pm 130$~\kms.  Using the relationships between \vsi\ and
maximum-light velocity, \vsiz\ \citep{Foley11:vgrad}, we derive
$v_{\rm Si~II}^{0} = -10,700 \pm 300$~\kms; therefore, \vsiz\ of
SN~2011iv is \about 800~\kms\ lower than the median value for the
\citet{Foley11:vgrad} sample (it has a lower velocity than 84\% of all
SNe~Ia in the sample).

\subsection{Comparison to Other Supernovae}

\begin{figure*}
\begin{center}
\epsscale{1.1}
\rotatebox{0}{
\plotone{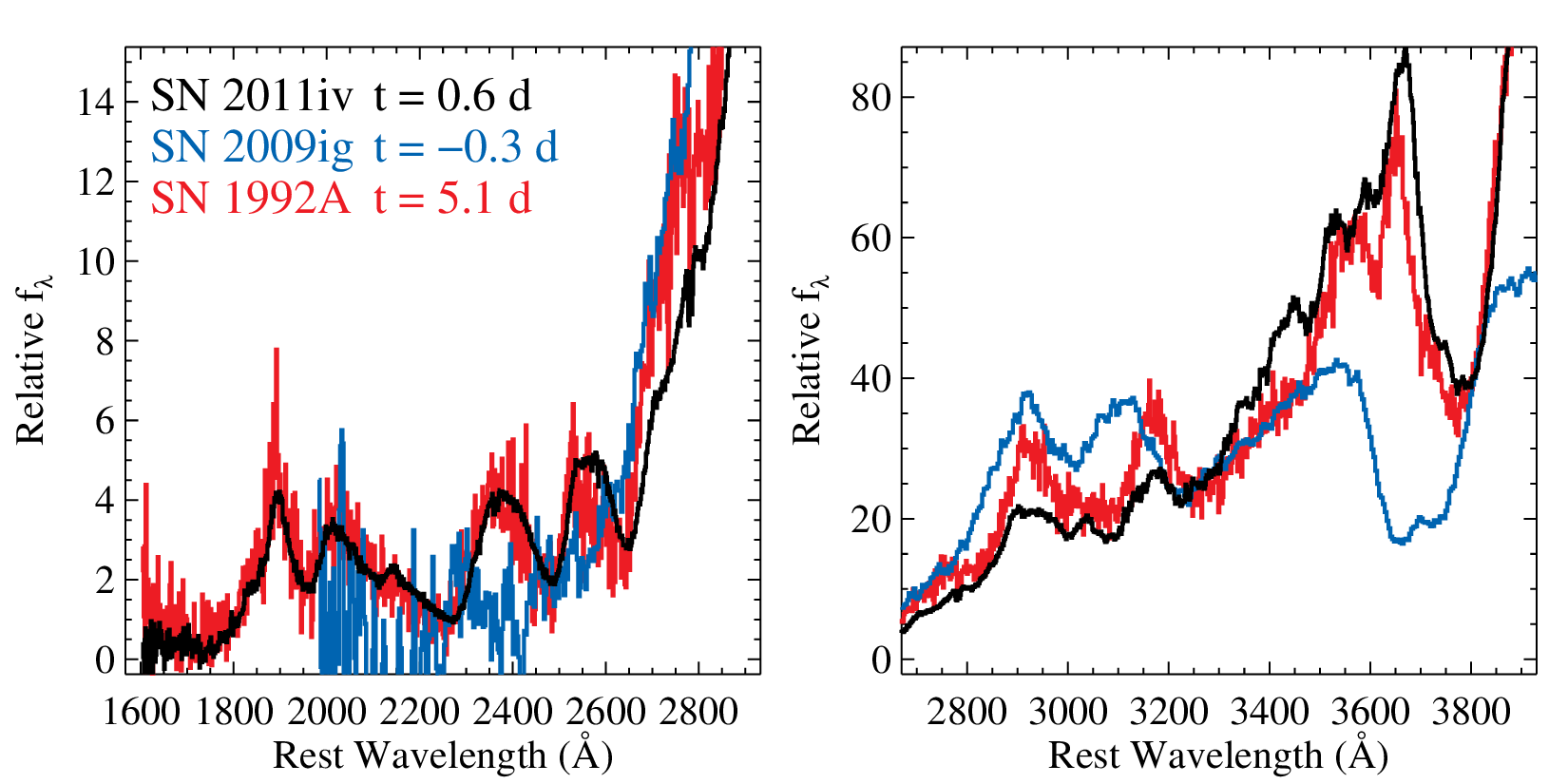}}
\caption{Near-maximum-light spectra of SNe~2011iv (black), 1992A (red;
\citetalias{Kirshner93}), and 2009ig (blue; \citealt{Foley12:09ig}).
The SN~2009ig spectrum is the combination of the $t = -2.1$ and 1.5~d
spectra (to increase the S/N).  The spectra have been scaled to have
similar flux at 3250~\AA.}\label{f:uv}
\end{center}
\end{figure*}

In Figure~\ref{f:uv}, we compare our SN~2011iv spectrum to the
near-maximum UV spectra of SNe~1992A \citepalias{Kirshner93} and
2009ig \citep{Foley12:09ig}, which have both been dereddened by their
Milky Way values of $E(B-V) = 0.018$ and 0.032~mag \citep{Schlegel98},
respectively.  At $\lambda < 3000$~\AA, the features in the SN~1992A
and SN~2011iv spectra are remarkably similar, showing only minor
differences.  There is a slight wavelength offset in the features at
2300--2700~\AA, with the features being more blueshifted in the
SN~1992A spectrum.  Although SN~1992A was observed at an epoch
4.5~days later than SN~2011iv, SN~1992A had higher \vsiz, and the
higher ejecta velocity is likely the cause for the small differences
in the UV.  SN~2009ig is significantly different from the other two
SNe; its spectrum lacks the prominent features at 2380 and 2560~\AA\
seen in both SN~1992A and SN~2011iv.  There is perhaps an indication
that the features are blueshifted even further than in SN~1992A, but
the S/N is too low to be conclusive.  Similarly, the SN~2009ig
spectrum does not have a sufficiently high S/N at $\lambda <
2300$~\AA\ to say anything definitive about features in the far UV.

Examining the three SNe in the near UV (2800--3800~\AA), there is even
more diversity in their spectra.  The spectra of SNe~1992A and 2011iv
are still similar in this region, but SN~1992A has more pronounced
features, especially surrounding \ion{Fe}{2} $\lambda 3250$ that is
observed at \about 3000~\AA.  Again, SN~2009ig has a very different
spectrum from the other two SNe.  Besides the clearly different Ca
H\&K line profile, the biggest difference is in the continuum shape;
SN~2009ig is relatively flat in the near-UV, while the other two SNe
are much steeper.  SN~2009ig has a very large \vsiz, and the
interpretation of \citet{Foley11:vel} is that the large velocity range
for absorbing lines, which correlates with \vsiz, causes more line
blanketing in the near-UV.  Additional line blanketing, either from
velocity or differences in the ejecta composition, may also explain
the relatively flat spectral slope in the near-UV.  Similar to what
was seen with extremely low-resolution UV spectra \citep{Wang12}, this
sample of SNe~Ia appears to have a large diversity in near-UV
continua.

Over 2000--2500~\AA, SN~2011iv has a S/N of 28 per 1.5~\AA\ pixel,
while SN~1992A has a S/N of 2.6 per 2~\AA\ pixel.  The SN~2011iv
spectrum has a S/N that is \about 15 times that of the SN~1992A
spectrum.  This exquisite spectrum shows additional features that are
presumably hidden in the noise of the SN~1992A spectrum.
Specifically, there is an absorption feature at \about 2120~\AA\ in
the SN~2011iv spectrum that is not seen in the SN~1992A spectrum.

\subsection{Comparison to Models}

The majority of the SN luminosity is emitted at optical
wavelengths. Most of this flux is redistributed from shorter
wavelengths where IGEs provide significant line opacity due to
millions of atomic line transitions. Although optical spectra are not
particularly good at differentiating between models (R\"{o}pke et~al.,
submitted), all models must reproduce the main optical signatures of
SNe~Ia to be credible.

It has been proposed that UV spectra of SNe~Ia are a good tool to
constrain different explosion models and progenitor metallicities
\citep[e.g.,][]{Hoflich98, Lentz01}.  Our unique spectrum provides the
possibility to test this assertion.  For that purpose we compare two
different explosion models to our data.  Specifically, we take the 1D
deflagration model W7 \citep{Nomoto84:w7} and the state-of-the-art 3D
delayed-detonation model N100 (R\"{o}pke et~al., submitted).  Both
models yield \about $0.6 M_{\sun}$ of $^{56}$Ni, which is typical for
normal SNe~Ia \citep{Stritzinger06}.  For both models we obtained a
spectral time sequence with the Monte Carlo radiative transfer code
ARTIS \citep{Kromer09}.

To study the influence of progenitor metallicity, we created an
additional model W7$_{Z_{\sun}}$, in which we polluted the outer
layers of W7 (zones where the sum of the IGE abundances was lower than
solar IGE content, i.e., $v < -12$,500~\kms) with solar abundances.  A
comparison of maximum-light spectra of our three models and the UVOIR
spectrum of SN~2011iv is shown in Figure~\ref{f:model}.  Both the N100
and W7$_{Z_{\sun}}$ models have similar SEDs to that of SN~2011iv from
the UV to the NIR.  In contrast, the original W7 model has excess UV
flux.

The differences in the W7 spectra are mostly limited to the far-UV
($\lambda < 2800$~\AA), indicating that only this wavelength regime is
strongly sensitive to progenitor metallicity (see also
\citealt{Lentz01}).  If we want to use SNe~Ia as distance indicators
at $z \gtrsim 1.5$, for which the observed optical data correspond to
the rest-frame far-UV, light-curve fitting algorithms have to account
for a range of different progenitor metallicities.

In contrast, our two metallicity points for W7 indicate that in the
near-UV ($2800 < \lambda < 3800$~\AA), the influence of different
progenitor metallicities is weaker.  This region might be better
suited to differentiate between explosion models.  Indeed, the N100
delayed-detonation model shows different characteristics than W7 in
this regime and reproduces the near-UV features of SN~2011iv somewhat
better.


\section{Discussion \& Conclusions}\label{s:conc}

We presented a UVOIR spectrum of SN~2011iv, a relatively normal SN~Ia.
These data were obtained 0.6~days after $B$-band maximum light with
\textit{HST}/STIS and Magellan/FIRE.  This spectrum is the first
contemporaneous and continuous UVOIR spectrum of a SN~Ia.  It is also
the first published high-S/N, near-maximum, true-UV SN~Ia spectrum
since that of SN~1992A \citepalias{Kirshner93} and the earliest
high-S/N UV SN~Ia spectrum yet published.

We compared the UV spectrum of SN~2011iv to the near-maximum light UV
spectra of SNe~1992A and 2009ig.  SNe~1992A and 2011iv have very
similar UV spectra.  However, SN~2009ig has a spectrum that is
significantly different from that of the other two SNe.  Specifically,
SN~2009ig lacks two prominent features at 2300--2700~\AA\ (though
this may be the result of its higher ejecta velocity), and it has a
relatively flat near-UV continuum.

Although the specific reasons for the differences between the UV
spectra of the three SNe are not yet clear, the SNe have several
additional observational differences.  SNe~1992A, 2009ig, and 2011iv
have S0, Sa, and E1 host galaxies; $\Delta m_{15} (B) = 1.48$, 0.89,
and 1.69~mag; and $v_{\rm Si~II}^{0} = -12$,900, $-13$,500, and
$-10,$700~\kms, respectively \citep{Foley11:vgrad, Foley12:09ig}.  All
three SNe have early-type host galaxies, so there is no clear
difference in progenitor age or metallicity; additional SNe must be
observed to probe the effects of environment on SN~Ia UV spectra.
Nonetheless, \citet{Sauer08} were able to produce different near-UV
continua (qualitatively similar to what is seen in Figure~\ref{f:uv})
by varying the amount of Ti and Cr generated in the explosion.
Specifically, models with less Ti/Cr had flatter near-UV continua.

SNe~1992A and 2009ig have similar \vsiz, while SN~2011iv has
significantly lower \vsiz; however, SNe~1992A and 2009ig have
different UV spectra, while SNe~1992A and 2011iv have very similar UV
spectra.  Finally, SNe~1992A and 2011iv are relatively fast-declining
SNe~Ia, while SN~2009ig is a relatively slow-declining SN~Ia.  It
appears that the characteristics of SN~Ia UV spectra are better
predicted by light-curve shape than ejecta velocity.  We hypothesize
that $^{56}$Ni generation, which regulates the decline rate
\citep{Kasen07:wlr}, is dominant over the kinetic energy per unit mass
in the formation of the UV spectra of SNe~Ia near maximum light.

UV spectra are particularly sensitive to progenitor metallicity and
the details of the explosion.  The maximum-light spectrum of SN~2011iv
is similar to both the delayed-detonation N100 model and the
solar-metallicity polluted W7$_{Z_{\sun}}$ model spectra; the standard
W7 model has significantly more far-UV flux than SN~2011iv.
Additional data will be useful in determining the full density
structure of the explosion.

This first spectrum shows the power of UV spectroscopy for
understanding the progenitors and explosions of SNe~Ia.  Our full 
set of UV, optical, and NIR data for SN~2011iv, including seven
\textit{HST} spectra, will be analyzed in a later paper.
SN~2011iv is one of the best-observed SNe, and detailed analysis of
this full dataset should provide even stronger constraints on
explosion models.

\begin{acknowledgments} 

\textit{Facilities:}
\facility{HST(STIS), Magellan:Baade(FIRE), PROMPT}

\bigskip
This research was supported by a Clay Fellowship (R.J.F.), an NSF
Graduate Research Fellowship (E.R.N.), the TABASGO Foundation
(A.V.F.), and NASA/{\it HST} grant GO-12592. G.P.\ acknowledges
support by the Proyecto FONDECYT 11090421, proyecto regular UNAB
DI-28-11/R, and by the grant ICM P10-064-F (Millennium Center for
Supernova Science), with input from ``Fondo de Innovaci\'{p}n para la
Competitividad, del Ministerio de Econom\'{i}a, Fomento y Turismo de
Chile.''  S.T.\ is supported by the DFG (Transregional Collaborative
Research Center TRR~33). The simulations were performed at JSC, 
J\"{u}lich, Germany (grants PRACE042, HMU14/20).

\end{acknowledgments}

\bibliographystyle{fapj}
\bibliography{../astro_refs}


\end{document}